\documentclass[12pt,peerreview,draftclsnofoot,letterpaper,oneside,onecolumn]{IEEEtran}

\usepackage{amsfonts,amsmath,amssymb}
\usepackage{dsfont}
\usepackage{cite}
\usepackage{graphicx}
\usepackage{booktabs}

\usepackage{xcolor}

\usepackage{caption}
\usepackage{bm}
\usepackage{bbding}
\usepackage{amsmath}
\usepackage{enumerate}
\usepackage{multirow}
\usepackage[ruled]{algorithm2e}
\usepackage{makecell}
\usepackage{subcaption}
\usepackage{array}

\usepackage{graphicx}

\usepackage{url}
\usepackage{amsthm}

\renewcommand{\IEEEQED}{\IEEEQEDopen}

  \newcommand{\figref}[1]{\figurename~\ref{#1}}

\begin{document}
\title{Intelligent Reflection as a Service (IRaaS): System Architecture, Enabling Technologies, and Deployment Strategy}
\author{Wei~Wang,~\IEEEmembership{Senior Member,~IEEE}, and Yutian~Shen
\thanks{

W. Wang is with the School of Information Science and Technology, Harbin Institute of Technology, Shenzhen 518055, China (e-mail: wang\_wei@hit.edu.cn)

Y. Shen was with the Business School of Shenzhen Technology University, Shenzhen 518118, China (e-mail: harryyutianshen@gmail.com).
}
}
\maketitle

\begin{abstract}
Reflecting intelligent surface (RIS) is a promising technology for 6G mobile communications. However, identifying the niche of RIS within the mobile networks is a challenging task. To mitigate the escalating system complexity of mobile networks, we propose the concept of Intelligent Reflection as a Service (IRaaS), and discuss its system architecture, enabling technologies, and deployment strategy, respectively. By leveraging technologies such as resource pooling, service-based architecture (SBA), cloud infrastructure, and model-free signal processing, IRaaS empowers telecom operators to deliver on-demand intelligent reflection services without a radical update of current communication protocols.  In addition, IRaaS brings a novel deployment strategy that creates new opportunities for the vendors of intelligent reflection service and balances the interests of both telecom operators and property owners. IRaaS is expected to speed up the rollout of RIS from both technical perspective and commercial perspective, fostering an authentic smart radio environment for future mobile communications.
\end{abstract}

\section{Introduction}

The low-cost passive programmable metasurfaces are capable of tailoring electromagnetic wave and are thus envisioned as a revolutionary technology that changes the paradigm of wireless communications from ``adapting to wireless channels" to ``changing wireless channel" \cite{basar2019wireless}. When applied to wireless channel reconfigurations, programmable metasurfaces are usually termed as reflecting intelligent surfaces (RISs) or intelligent reflecting surfaces (IRSs). Both IRS and RIS are characterized by intelligence, which refers to the capability of  effectively configuring the reflecting elements of RIS for a better radio propagation environment.  Compared with the conventional active wireless devices, RIS incurs no additional power consumption and is free from the thermal noise introduced by radio frequency (RF) modules, but the cost is its inability to sense signal. 

A significant challenge that telecom operators are confronted with the massive deployment of RIS is how to smoothly shift from current radio infrastructure and mobile standards to RIS-assisted wireless communications, given the former’s intricate system structure and complex workflow. Following the conventional practice of MIMO communications, most solutions of RIS configuration \cite{CEBF} depend on the availability of channel state information (CSI) of RIS-Tx, RIS-Rx, Tx-Rx sub-channels. The acquisition of the sub-channel CSIs hinges upon the coordination among Tx, Rx, and RISs, thereby creating a strong coupling between the wireless communication system and the RISs. The coupling inevitably diminishes the flexibility, scalability, and manageability of the RIS-assisted wireless communication systems. Furthermore, a radical update of the existing wireless protocols, {which were originally designed only for Tx–Rx operation}, is imperative to accommodate {the  coordination of Tx, Rx and RISs}.

{In addition, public might be against the massive deployment of RISs due to aesthetics and health concerns}. Note that RIS can be classified into two categories: the non-transparent RIS and the transparent RIS \cite{etsiTransRIS}. Non-transparent RISs are anticipated to be installed on building facades and interior walls to improve outdoor and indoor radio environments. Transparent RISs  are designed for installation as windows of buildings to enhance both outdoor and indoor radio environment. Comparable to the hurdles encountered in deploying base stations, the installation of RISs in residential areas, despite their passive nature, faces inherent opposition from the public. The opposition to the installation of RISs stems from multiple factors such as cost, aesthetics, and health concerns. Persuading property owners to opt for transparent RISs, despite their higher costs compared to conventional glass windows, requires an effective strategy. Additionally, securing permission for the installation of large-scale non-transparent RISs on residential building facades necessitates addressing concerns about their visual impact. 

{On the other hand, the mobile network has become more dynamic and flexible, enabled by the service based architecture (SBA) and characterized by the extensive use of cloudified network functions (NFs) of the 5G core network (5GC) \cite{zeydan2022service}.} The 5GC with SBA consists of a set of interconnected NFs that are softwarized and cloud-native. The NFs access the services of others through the authentication and authorization mechanism. Hence, each individual service can be autonomously updated with minimal disruption to other services, facilitating on-demand deployment and enabling automation and agile operational processes.  In the meantime, the telecom industry organizations, e.g., O-RAN alliance, are also working towards the virtualization of radio access network (RAN) to introduce SBA to RANs, aiming to increase flexibility and scalability, reduce costs, and meet growing demand \cite{zeydan2022service}. 

{How to enable effective integration and interaction between RIS and RAN has been a key challenge for RIS applications and has been widely discussed  \cite{RISintegration, PMvirtual}. Different from existing solutions, we propose the concept of Intelligent Reflection as a Service (IRaaS), which offers intelligent reflection capabilities in an on-demand manner and enables an incremental evolution of RIS deployment.}   In this article, we discuss the challenges and opportunities of IRaaS, aimed at expediting the widespread implementation of RISs. Exploiting the technologies such as resource pooling, SBA, cloud infrastructure, and model-free signal processing, IRaaS enables telecom operators to provide on-demand services of intelligent reflection without a radical update of communication protocols. In addition, IRaaS brings a new business model that opens new opportunities for the vendors of intelligent reflection service and balances the interests of both telecom operators and property owners. IRaaS is expected to accelerate the massive deployment of RIS, fostering  an authentic smart radio environment for the future mobile communications.

\section{The Status Quo of RIS Configurations}

From the perspective of model dependance, RIS configurations can be categorized into two types, namely model-based RIS configurations and model-free RIS configurations. This section reviews both configurations, as they critically influence the architectural design of RIS-assisted mobile networks.

\subsection{Model-based RIS configurations}

Model-based RIS configurations treat the RIS as a part of the wireless communication system, and they rely on the awareness of sub-channel status \cite{basar2019wireless, mei2022intelligent}. To acquire sub-channel CSI, conventional channel estimation techniques must be updated to introduce coordination among the transmitter, receiver and RIS to compensate for the RIS's inability to sense. Coordinated channel estimation aims to acquire the CSIs of the Tx-Rx sub-channel, Tx-RIS sub-channel, and Rx-RIS sub-channel, and then optimize the reflection pattern of the RIS based on these sub-channel CSIs. However, channel estimation incurs excessive training overhead and is highly dependent on accurate reflection parameters, which are difficult to control with wideband signals. Additionally, the decoupled structure of RISs from the radio access network (RAN) results in poor scalability and high maintenance costs.
 
\subsection{Model-free RIS Configurations}

Model-free RIS configurations treat the RIS as an integral part of the radio environment rather than as a component of the wireless communication system \cite{ModelFree}. These configurations determine the passive reflection pattern of the RIS directly from channel measurements and performance indicators of the wireless communication system, without inferring the sub-channel CSI. Model-free RIS configurations aim to maximize the accumulated or expected overall performance through the continuous small-scale adjustment of reflection coefficients. The goal of model-free methods is to decouple the RIS system from the wireless system, creating a scenario where wireless communication operates seamlessly, unaware of the RIS system. The RIS, in turn, configures itself independently using only the information provided by interfaces connected to the wireless communication system.

\section{The Architecture of IRaaS}

IRaaS refers to the provision of intelligent reflection  services via the RIS resource pool and the cloud computing infrastructure. Analogous to other ``as a service" offerings such as Platform as a Service (PaaS), Software as a Service (SaaS), and Artificial Intelligence as a Service (AIaaS), IRaaS provides both telecom operators and mobile users with the ability to improve link qualities without the necessity of infrastructure investments or specialized expertise. Instead of embedding RISs into the RAN as a whole, the RISs are decoupled with the RAN in the proposed framework.  This implies that the RIS resource pool functions as an autonomous layer, interacting with both the RAN and core network (CN) through interfaces.



\begin{figure}[htp]{
\begin{center}{\includegraphics[width=1\textwidth]{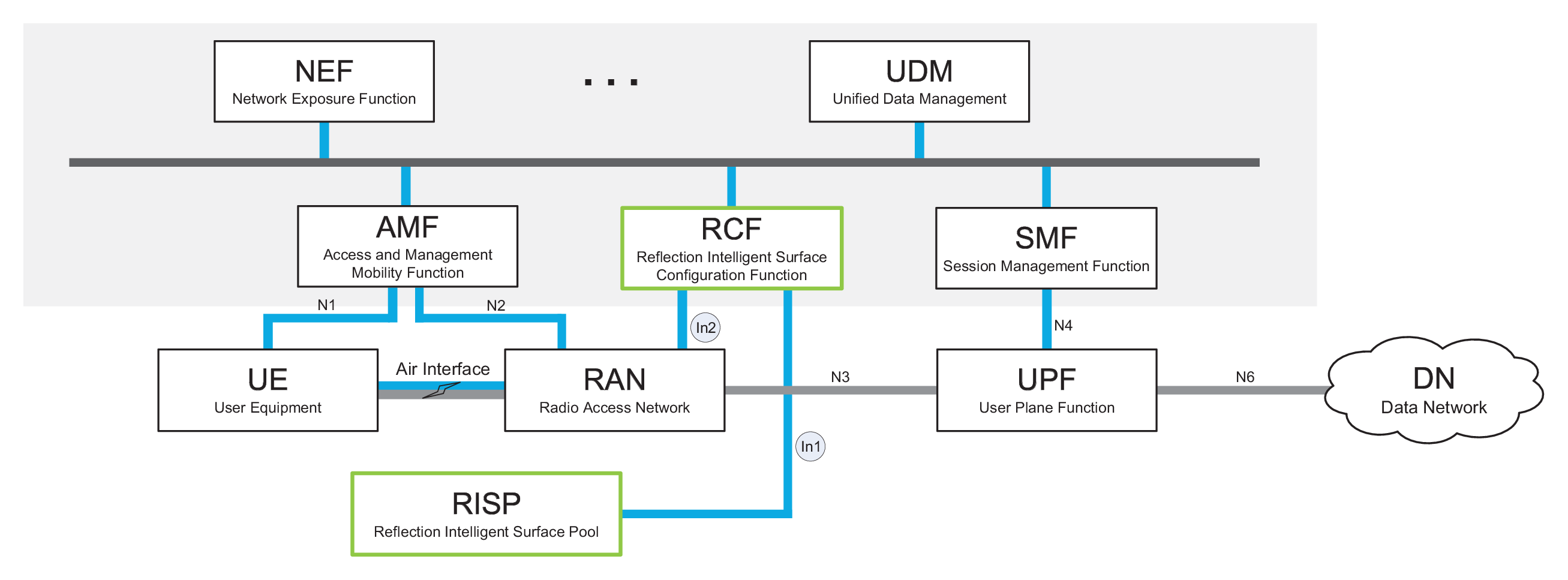}}
\caption{The architecture of 5G mobile network with IRaaS, where RCF and RISP are newly defined entities, and ln1 and ln2 are newly defined interface.} \label{SBA1}
\end{center}}
\end{figure}

{ 
\subsection{Design Philosophies}

Our view is that RIS deployment should preferably follow the philosophies such as 
\begin{itemize}
\item
{\bf Incremental evolution:} The RIS system should expand gradually, without requiring a fully built-out infrastructure from the beginning. 
\item
{\bf Best-effort principle:} The configuration of RIS follows an Internet-like philosophy, where the RIS system delivers reflection services without guaranteed Quality of Service (QoS), leaving strict performance assurances to upper-layer scheduling or the RAN.
\item
{\bf Loose coupling:} IRaaS connects RAN using standard protocols, avoiding the enormous complexity of unified management.
\end{itemize}
The philosophies of incremental evolution, the best-effort principle, and loose-coupling collectively help reduce the complexity and enhance the scalability of next-generation RAN. They also facilitate lowering both the cost and the deployment risks associated with RIS deployment.

}

\subsection{The Architecture of 5G NR with IRaaS}

\subsubsection{{5GC SBA}} 
Without loss of generality, we use 5G NR to demonstrate the integration of IRaaS into mobile networks. In the SBA of 5GC, NFs interact by providing and consuming services via well-defined interfaces, as shown in \figref{SBA1}.   The NG-RAN, i.e., the RAN of 5G NR, which comprises a set of gNBs, i.e., 5G base stations, is connected to the 5GC, via the NG interface. There are two interfaces under the NG interface, i.e., NG-C (a.k.a., N2 interface) for the control plane that connects with the Access and Mobility Management Function (AMF), and NG-U (a.k.a., N3 interface) for the user plane that connects with the User Plane Function (UPF) \cite{gpp38413}. The user equipment (UE) is connected to AMF through the N1 interface, and N1 is a logical interface, which is indeed the combined path from UE to RAN and onward to AMF. The N4 interface connects the UPF to the Session Management Function (SMF) and caters to the key session management procedures. The N6 interface is used to connect the UPF to the Data Network (DN).

\subsubsection{{New Entities and Interfaces of IRaaS}}
IRaaS is realized through two crucial entities, namely the RIS pool (RISP) and the RIS configuration function (RCF), where the former provides the hardware platform of electromagnetic (EM) reflection, and the latter provides intellectual support to maximize the capability of RISs in optimizing the radio environment. The intelligent reflection service is interconnected with the 5GC via interfaces. Among the interfaces, we designate the In1 interface to connect the RISP and the RCF such that the configuration settings are received by the RISP from the RCF. Additionally,  the In2 interface is designated to support information exchange between the RAN and the RCF.

{
RCF, the control plane of IRaaS,  can be  implemented as either an application function (AF) or an NF. This choice represents a trade-off in the level of decoupling. An AF implementation is preferable for a higher degree of decoupling, whereas an NF implementation offers a more tightly integrated approach. 
\begin{itemize}
\item 
{\bf AF:} From an architectural decoupling perspective, implementing the RCF as an AF is simpler and avoids added complexity or misalignment with 3GPP specifications. As an AF, the RCF can only influence the network indirectly via NEF, and  does not involve core network control.
\item
{\bf NF:} If implemented as an NF, the RCF would be more tightly integrated with the CN, thereby enabling more direct control, lower latency, and unified security and management.
\end{itemize}
If the RCF is owned by an affiliated subsidiary of the telecom operator, then deploying the RCF as an NF is feasible. In contrast, if the provider is an external entity, the RCF can only be positioned as an AF, as operators are typically reluctant to open their core network to non-internal parties. However, we remain certain that the RCF should be decoupled from the RAN.}

{Note that the RCF can interact with 5G core network functions, e.g., the AMF, SMF, and UDM, in two feasible ways, depending on the deployment scenario and coupling level. Both approaches are aligned with the 3GPP SBA. As an NF, the RCF can be implemented as a new native network function that directly participates in the SBA. Alternatively, as an AF, the RCF indirectly invokes services from the AMF, SMF, and UDM through the NEF, using their standardized APIs.}



\subsection{The Workflow of IRaaS}

The RCF of IRaaS primarily performs the following functions:
\begin{itemize}
\item \textbf{RIS registration}: The RCF is responsible for registering a RIS with the mobile network, and storing in the registry the RIS’s key information essential for RIS configuration, e.g., location, orientation, size of the reflective element array, phase quantization level, and etc. 
\item {\textbf{Session management}: The RCF manages the establishment, modification, and termination of reflection service according to the requests from gNB.  It interacts with other NFs to obtain the actual performance metrics of the QoS Flows within Packet Data Unit (PDU) sessions, such as throughput and packet loss. These QoS indicators enable the RCF to determine whether the current RIS allocation and configuration satisfy the QoS requirements.}
\item \textbf{RIS allocation}: The RIS resource pool serves multiple UEs with provisional and scalable intelligent reflection  services. Those services can be dynamically adjusted based on the demand from UEs without any discernible changes to the UEs. The RCF is responsible for RIS allocation, and ensures that the RIS, as a type of resource, is appropriately allocated into the UEs that subscribe to the intelligent reflection service. 
\item \textbf{Radio environment analysis and RIS coefficient derivation}: The RCF collects and archives radio environment data, e.g., CSI and link quality indicator, from the RAN. Subsequently, it conducts comprehensive analysis leveraging model-free algorithms, e.g., deep learning, reinforcement learning and derivative-free optimization. Following this analysis, the RCF governs the phase shift adjustments of the allocated RIS units. 
\end{itemize}
 
\begin{figure}[htp]{
\begin{center}{\includegraphics[width=0.8\textwidth]{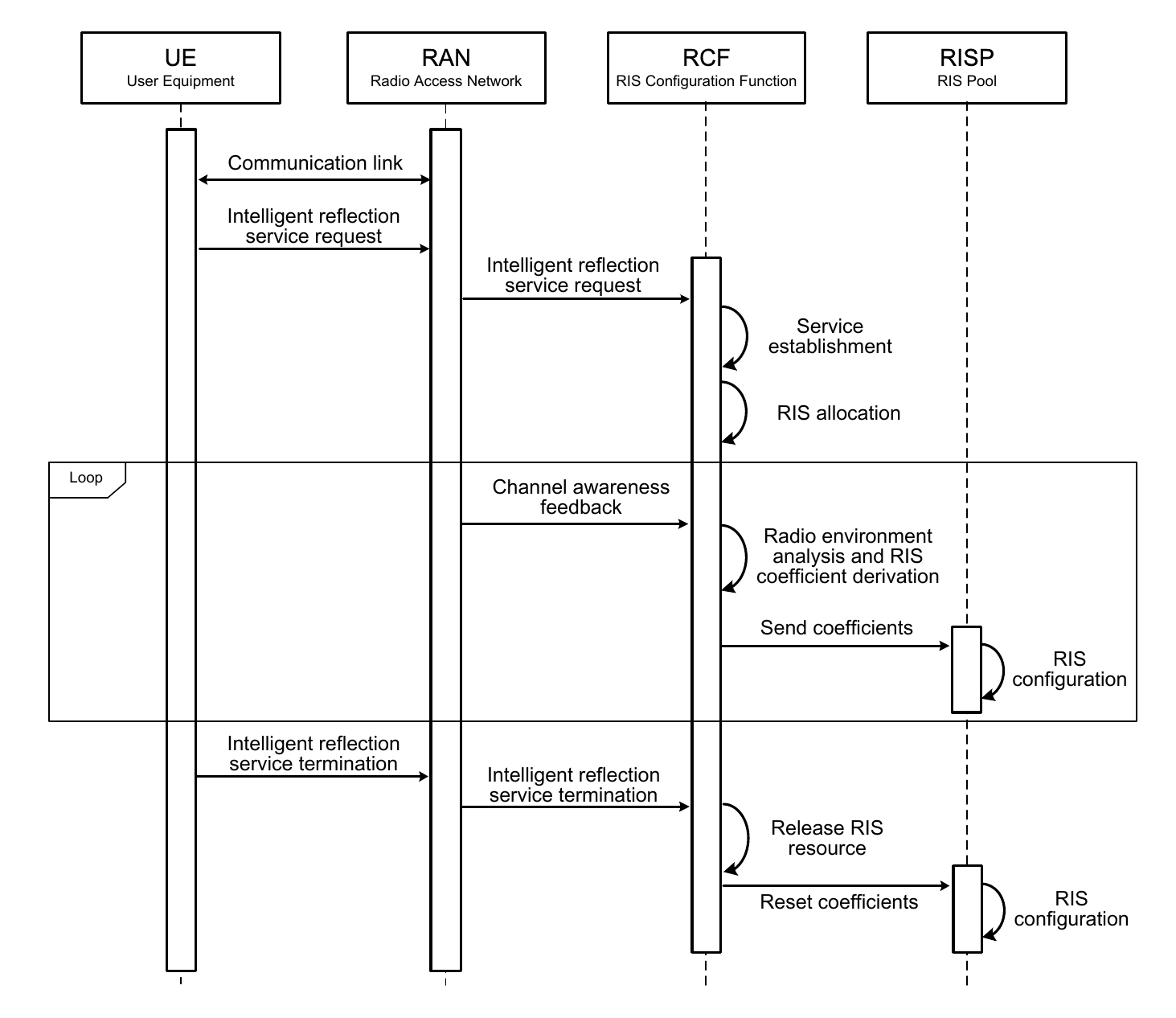}}
\caption{The sequence diagram of IRaaS} \label{SequenceFlow}
\end{center}}
\end{figure}

To explain the workflow of IRaaS,  we use a sequence diagram, shown in \figref{SequenceFlow}, to  illustrate how RCF and RISP interact with the mobile network by depicting the sequence of messages exchanged between them.  IRaaS, as a service within the communication processes, is activated selectively to enhance communication quality when necessary. For instance, it becomes operational when the UE experiences high packet loss rates caused by signal blockages. Typically, the initiation of IRaaS stems from the UE, as it possesses the most direct awareness of its link quality. The process begins with the UE sending a service request to RCF through the air interface and the In2 interface. Upon receipt of this request, RCF proceeds to allocate RIS resources based on the disparity between the UE's current communication quality and the desired quality level. Subsequently, utilizing the allocated RIS resources, model-free signal processing algorithms are employed to optimize the reflection coefficients based on radio environment information. The derived intermediate coefficients are then transmitted to the RIS via the In1 interface to update  the RIS's impact on the radio environment. 
The service terminates either when the UE's communication task concludes or when its requested communication quality degrades. Subsequently, the RCF releases the UE's requested RIS resources back to the pool and resets their coefficients.

\section{The Enabling Technologies of IRaaS}
\begin{figure}[htp]{
\begin{center}{\includegraphics[width=0.6\textwidth]{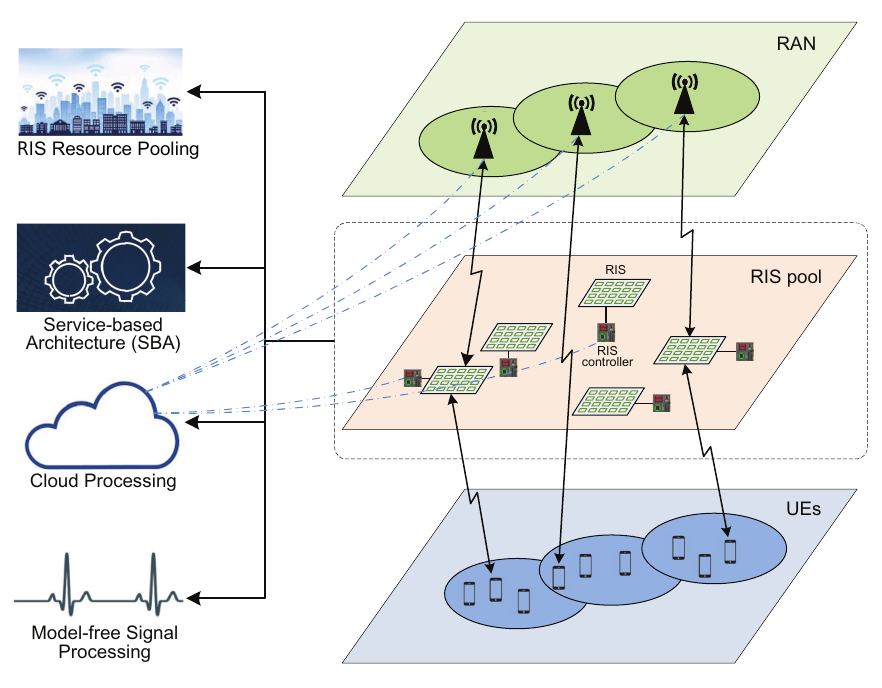}}
\caption{The enabling technologies of IRaaS and the logical relationship between RAN, UE and RIS } \label{IRSpool}
\end{center}}
\end{figure}

The enabling technologies of IRaaS and the logical relationship between RAN, UE, and RIS are presented in \figref{IRSpool}.

\subsection{RIS Resource Pooling}

Resource pooling is applied to aggregate multiple RISs to act like one powerful resource for the more flexible and more efficient utilization of RISs in the complex radio environment \cite{qadir2016resource}. The RIS resource pool consists of a set of RISs shown in \figref{IRSpool}.   Resource pooling grants the on-demand availability of RISs as a resource without direct management by the base stations (BSs) or the UEs. The RIS resource pool can be assigned to different tasks or shared by several tasks. For instance, a RIS in the resource pool might be allocated to enhance the overall communication quality of multiple users occupying different radio resources \cite{ModelFree}.

However, the resource pooling technology in wireless domain is fundamentally different from those used in cloud computing and faces great challenges. In cloud computing, the capacity of resources such as  CPU, GPU, memory, and storage, is largely independent of their physical locations, except for considerations related to latency. By contrast, the capacity of RIS's signal enhancement is strongly related to the distance between RIS and the transceivers, due to the propagation loss of electromagnetic wave. In other words, the resource map of RIS is distance-dependent and is related to the locations of Tx and Rx.

\subsection{Cloud Processing}

The efficient RIS resource allocation and effective RIS configuration depend on the computing power of the RIS configuration units. Considering the delay-tolerant but computationally demanding nature of RIS configurations, the RIS configuration units can be realized in the cloud, which allows for a much more powerful computing resource pool to achieve low-cost operation of the RISs. Analogous to the virtual baseband units  in the cloud radio access networks (C-RAN) \cite{C-RAN, zeydan2022service}, virtual {RCF} units can be realized in the centralized cloud located away from the RISs to leverage the computing power of the existing cloud infrastructure as shown in \figref{IRSpool}. 
{By leveraging centralized cloud computing with a global network view, IRaaS enables coordinated and globally optimized RIS reconfiguration \cite{shen2021holistic}. }

\subsection{Model-Free Signal Processing}

As a service, intelligent reflection should be disaggregated from the other functions of RAN. However,  the conventional model-based RIS configurations are strongly coupled with the RAN \cite{mei2022intelligent}. {Specifically, they require the acquisition of the CSI for the Tx–Rx, Tx–RIS and RIS–Rx sub-channels to enable RIS configurations.} To support IRaaS, it is necessary to adopt model-free signal processing techniques in the physical layer for decoupling \cite{ModelFree}. {These techniques treat the RIS as an independent component of the RAN and optimize RIS configurations based on the interactions between the RIS and the RAN. As an example, the RIS controller periodically applies candidate reflection configurations, collects performance feedback from the RAN, e.g., SINR or throughput, to sense the state of the radio environment, and refines the configuration through trial-and-feedback iterations, thereby realizing RIS optimization without explicit channel modeling.}  Within the realm of model-free signal processing techniques in RIS configuration, there exists a broad spectrum of approaches that operate independently of the sub-channel CSI that forms the aggregated wireless channel. This encompasses methodologies such as extremum seeking control (ESC), a.k.a., derivative-free optimization \cite{larson2019derivative}, alongside deep reinforcement learning \cite{wang2022intelligent}, and various other machine learning technologies. 

{Model-free signal processing techniques are particularly well suited to the SBA, as they are independent of sub-channel CSI and the internal working mechanisms of wireless communication systems. With model-free RIS configuration, the operation of the RIS only requires exchanging output information with the RAN through well-defined interfaces, without necessitating any modification to existing RAN functionalities.}

\subsection{{On-Demand Intelligent Reflection Service Enabled by SBA} }
{The intelligent reflection service provided by IRaaS is an on-demand service that can be activated only when needed.} For instance, the intelligent reflection service can be initialized by a user to construct a virtual direct link to substitute a blocked mmWave direct link, and it can remain deactivated when not needed.  {In this context, the intelligent reflection service is deployed without affecting the standalone operation of the 5G RAN. Such a design reflects the philosophies of incremental evolution, best-effort, and loose coupling. }

{The SBA offers the most natural and standardized mechanism for supporting on-demand intelligent reflection service in 5G and beyond. First, the SBA provides service registration and discovery, allowing a function to be activated only when another function explicitly requests its service. Second, its RESTful, API-driven interactions eliminate the need for persistent bindings between NFs, enabling a clean separation of lifecycle management and on-demand invocation. Third, SBA ensures interoperability and standardization across vendors, technically enabling IRaaS to be provided by a third party. }

\subsection{Overview of the Technological Advantages}

{In summary, from a technological perspective, IRaaS offers several key advantages: (1) architectural decoupling and backward compatibility, (2) high scalability, (3) global optimization capability, and (4) reduced signal processing complexity. Specifically, advantages (1) and (2) are enabled by RIS pooling and the SBA architecture, (3) is enabled by cloud processing, and (4) is achieved through model-free signal processing.}




\section{The Deployment Strategy of IRaaS}

\begin{figure}[htp]{
\begin{center}{\includegraphics[width=1\textwidth]{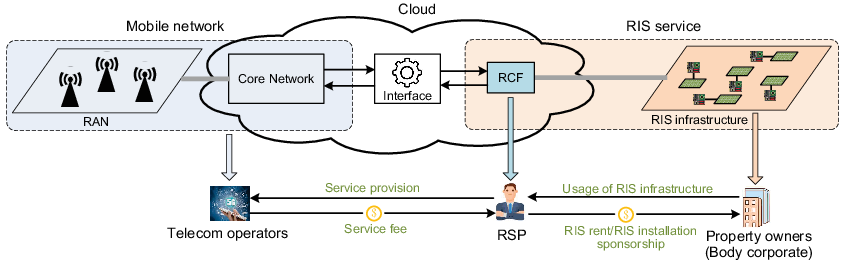}}
\caption{The three-party business model of IRaaS} \label{RISprovider}
\end{center}}
\end{figure}
 
The deployment of RIS system involves not only technical considerations but also various non-technical factors, such as commercial benefits, property-use agreements, and ownership structures. The business model of IRaaS plays a pivotal role in faciliating the deployment of RIS system within and beyond mobile networks, which precisely outlines the methods of income sources, ownership of the RIS infrastructure, required investment, and management of operations and maintenance. It serves as the blueprint for ensuring efficient functioning and sustainability of the RIS system in the evolving mobile network landscape.

\subsection{The Three-Party Business Model}
{In contrast to the traditional telecom paradigm, in which operators both build and operate the infrastructure, we propose a three-party business model for the deployment and management of RIS.} We propose to introduce a new actor, i.e., reflection service provider (RSP), as the intermediary, to facilitate the smooth and gradual evolution of RIS-assisted mobile networks in the business model shown in \figref{RISprovider}. The roles of the three parties in our proposed new business model are explained as follows.

(1) \textbf{RIS infrastructure owners:} The property owners, or the body cooperate, own the RIS infrastructure installed on their buildings. The implementation of RIS infrastructure inevitably encounters civic obstacles stemming from its installation on buildings. In addition,  the capital expenditure (CAPEX) costs to purchase and install the RIS system poses a significant burden on the construction of the next-generation mobile network, which will finally lead to escalated plan rates. Drawing inspiration from the solar industry, involving property owners as stakeholders in the RIS business can mitigate resident resistance and alleviate CAPEX pressures. Ways of incorporating property owners into the RIS business include installation sponsorship, a long-term rent contract and etc. In this regard, the property owners involve in the RIS business without extra upfront costs. 

(2) \textbf{Reflection service providers (RSPs):} RSP is a third party entity that acts as the bridge between RIS infrastructure owners and telecom operators. RSPs hold the usage rights of the RIS infrastructure and generate revenue through service fees charged to telecom operators for delivering intelligent reflection services.  RSPs are responsible for operating, maintaining, and ensuring optimal performance of the RIS system. The core technical competitiveness of RSPs is their RIS configuration algorithm that relies on signal processing and artificial intelligence (AI), and the core business competitiveness of RSPs is their cost control in massive RIS deployment. RSPs can mitigate investment risks by starting their RIS business from a small region, as the decoupled nature of the system allows for flexible deployment adjustments in terms of location and density at any time.

(3) \textbf{Telecom operators:} The telecom operators that own the mobile network are customers of the RIS business. They circumvent the upfront investment in RIS infrastructure through this new business model. Telecom operators procure reflection services from RSPs to tackle poor coverage issues, particularly in high-frequency bands like mmWave. In addition, telecom operators can offer on-demand intelligent reflection service to users willing to pay extra for improved signal quality. Telecom operators reap the benefits of an efficient RIS system by reducing the deployment cost of RAN. 
Nevertheless, it is essential for telecom operators to support RSPs during the start-up stages of the RIS business through various means.

\subsection{Benefits of the Business Model for RIS Deployment}

A good business model not only unlocks financial incentives and opportunities but also aligns with the industry's dedication to a sustainable and prosperous future for all stakeholders. The proposed business model is advantageous in the following aspects.

\begin{itemize}
\item {A win–win empowerment model for property owners:} The proposed business model provides property owners with  the chance to make profit from their building ownership, while simultaneously providing property owners, who are mobile network users in the meantime, with improved communication quality at a reduced rate.
\item {A high-barrier, asset-exclusive business paradigm for RSPs:} The proposed business model introduces new avenues for wealth generation and establishes a novel business entity, known as RSP. RSPs operate their business leveraging the cutting-edge technologies, e.g., information metamaterials and AI, which come with a high technical barrier to entry. Moreover, the rental agreement with RIS owners grants RSPs a form of resource monopoly, ensuring stable and sustained revenue.  
\item {A zero-CAPEX, backward-compatible smart radio environment solution for telecom operators:} The proposed business model enables telecom operators to achieve a smart radio environment with zero upfront costs, thereby mitigating their investment risks in next-generation mobile networks. Furthermore, by offering an open interface, current mobile networks can seamlessly adopt the smart radio environment without requiring any updates. 
\end{itemize}

{
\subsection{Overview of IRaaS Components and Their Ownership}

Based on the above discussions, we use Table \ref{Table1} to summarize the key entities, interfaces, and their ownership. Overall, IRaaS introduces only minimal architectural changes to the RAN. Through an on-demand service model enabled by open interfaces, together with the three-party business model, it provides a practical approach to addressing technical and deployment challenges for RIS.
\begin{table}[htp]
\caption{A summary of the newly defined entities and interfaces of IRaaS}
\label{Table1}
\centering
\begin{tabular}{c c c}
\hline
\textbf{Entities/Interfaces} & \textbf{Function} & \textbf{Ownership} \\ \hline
RCF & RIS configuration (software) & RSP   \\
RISP & RIS pool (hardware) &  \makecell{ Owned by property owners\\ and leased by RSP} \\
Interface In1 &  \makecell{ The interface between RISP and \\  RCF for coefficient transmission }  & RSP \\ 
Interface In2 &  \makecell{  The interface between RAN and  \\ RCF for channel awareness feedback} & Telecom operators  \\\hline
\end{tabular}
\end{table}

}

\section{An Illustrative Example}
In this section, we use an example to illustrate the working mechanism of IRaaS, {and to show why it is technologically advantageous.}

\subsection{{Simulation Settings}}

Assume that four {UEs}, each equipped with $4$ antennas are served by a BS equipped with $16$ antennas within a cell. An optically transparent RIS, integrated into window glass and owned by the property owner, is managed by the RSP. The configuration function of the RIS, namely RCF, is realized remotely in the cloud. The UEs may subscribe to the service of intelligent reflections to their telecom operators through the core network, which is interconnected with the RCF.

\begin{figure}[htp]{
\begin{center}{\includegraphics[width=0.8\textwidth]{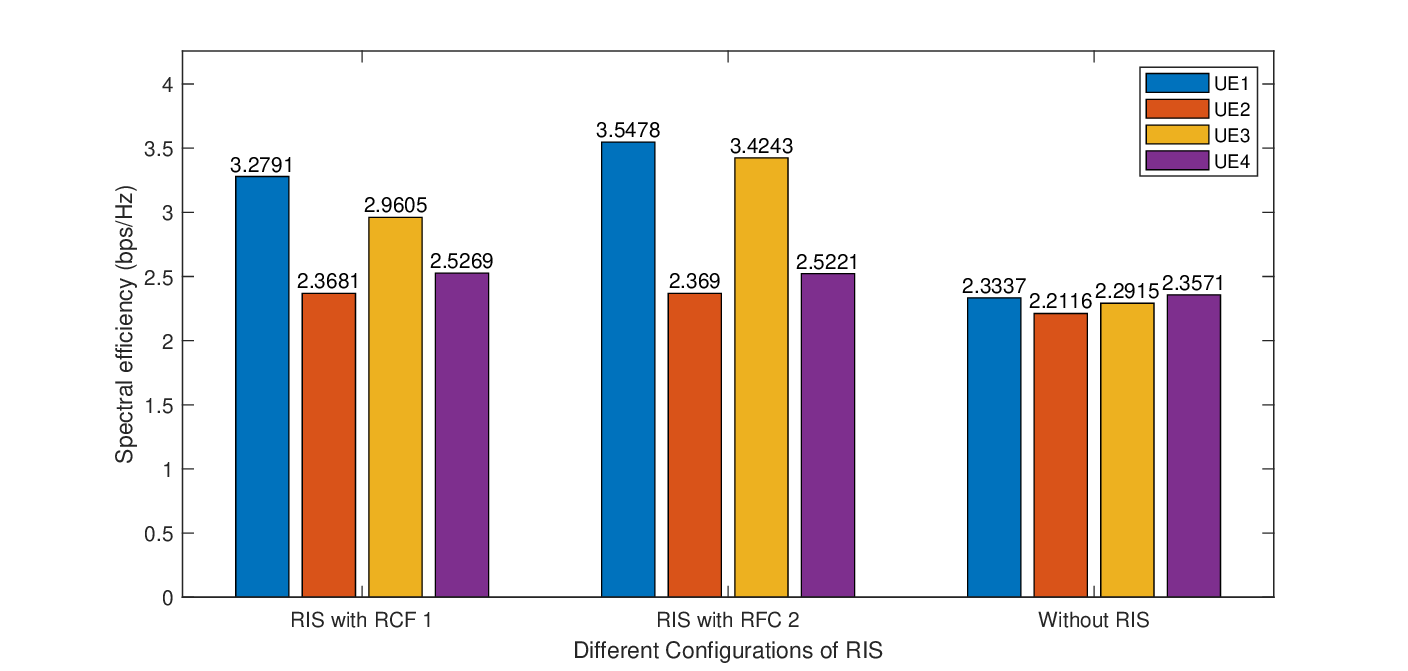}}
\caption{Communication performance under different configurations of RIS, where UE 1 and UE 3 subscribe to the service of intelligent reflection, while UE 2 and UE 4 do not.} \label{Performance}
\end{center}}
\end{figure}

The simulation setting is as follows. We adopt the practical model of RIS \cite{abeywickrama2020intelligent}, where the reflection coefficient is given by
$v_n = \frac{Z_n(C_n, R_n) - Z_0 }{Z_n(C_n, R_n) + Z_0}$, which is a function of $C_n$ and $R_n$ and  
varies according to the operating frequency of the incident RF signal. The reflected electromagnetic waves can be manipulated in a controllable and programmable manner by varying $C_n$ and $R_n$. The impedance of air is $Z_0 = 377$ $\Omega$. According to \cite{abeywickrama2020intelligent}, the equivalent model of the $n$-th reflecting element is represented as a parallel resonant circuit, with its impedance given by $ Z_n(C_n, R_n) = \frac{j \omega L_1 (j\omega L_2 + \frac{1}{j\omega C_n} + R_n)}{j \omega L_1 + (j\omega L_2 + \frac{1}{j\omega C_n} + R_n)}$, where $L_1$, $L_2$, $C_n$, $R_n$, and $\omega$ denote the bottom layer inductance, top layer inductance, effective capacitance, effective resistance, and angular frequency of the incident signal, respectively. $C_n$ is varied from $0.47$ pF to $2.35$ pF, $L_1 = 2.5$ nH, $L_2 = 0.7$ nH, $\omega=  2\pi \times  2.4\times 10^9$. Each UE occupies $20$ MHz bandwidth protected by a $2$ MHz guard band. The number of reflecting elements is $128$, RIS location is $(0.5, 0, 3)$, and BS location is  $(0, 0, 3)$, the four UEs are evenly distributed along a semicircular area centered at $(0, 0, 0)$. The wireless channel consists of $1$ line-of-sight (LoS) path and $4$ non-line-of-sight (NLoS) paths. Blockage probabilities of the BS-UE link and the RIS-UE link are $0.3$, while that of the BS–RIS link is $0$, and blockage results in a propagation attenuation of $30$ dB.

\subsection{{Observations and Discussions}}
UE 1 and UE 3 subscribe to the intelligent reflection service, whereas UE 2 and UE 4 do not. The performance of all UEs are illustrated in \figref{Performance}, based on which we draw the following observations.

  \emph{\textbf{Observation 1}}: The performance of signal enhancement by RIS is determined by RCF. {In RCF 1, the RIS is partitioned into two halves, with the reflection coefficients of each half optimized independently to improve the performance of UE 1 and UE 3, respectively; In RCF 2,  the reflection coefficients are optimized to enhance the overall performance of UE 1 and UE 3.} In \figref{Performance}, we adopt the model-free signal processing algorithm ESC \cite{ModelFree}. From the figure, it can be observed that {RFC 2 outperforms RFC 1 for both UE 1 and UE 3. This is because RCF 2 coordinates all RISs to optimize the performance of all users, enabling a global view and conferring global optimization capability.} 
   

  \emph{\textbf{Observation 2}}: Even UEs that do not subscribe to the intelligent reflection service can benefit from the deployment of RIS.  Under both RFC 1 and RFC 2, UE 2 and UE 4, which do not subscribe to the service, achieve better performance compared to the case without RIS deployment. This is because rich scattering is essential for MIMO communications, even if the UEs that do not subscribe to the intelligent reflection service are not explicitly optimized. {This behavior aligns with the principle of incremental evolution, where new capabilities are introduced in an on-demand manner, delivering performance gains to subscribed UEs without degrading other UEs' performance.}

  \emph{\textbf{Observation 3}}:  Model-free signal processing significantly reduces the complexity of RIS configuration, enabling  RCF to be compatible with different types of RIS.  While the reflection coefficient varies with the operating frequency of the incident signal (e.g., in \figref{Performance}, UEs 1–4 operate at different frequencies and thus exhibit different reflection coefficients), its precise expression is unnecessary in model-free signal processing. The data-driven nature enables pervasive adaptability to different types of RIS, aligning well with the principles of AI and making it suitable for deployment on general-purpose cloud hardware. Given that sufficient runtime data of the RIS-assisted mobile communication system is provided, the RSP can continuously improve the performance of RIS to reach excellence. 

   \emph{\textbf{Observation 4}}: For RIS manufacturers, it is sufficient to provide a standardized interface that enables interoperability among the RIS, RAN, and CN, {thereby ensuring architectural decoupling, backward compatibility, and high scalability of IRaaS}.  The primary challenge lies in reducing the cost of RIS to make it an affordable option for property owners who wish to deploy a smart radio environment to ensure high-quality signal coverage.

\section{Conclusion and Future Work}
In this article, we introduce the concept of IRaaS and discuss its enabling technologies, workflow, and business model, aimed at accelerating the rollout of RISs in wireless communication networks. IRaaS enables telecom operators to provide on-demand services of intelligent reflection without requiring radical updates to the current communication protocols. Additionally, the decoupling of the RIS system and the RAN, facilitated by IRaaS, prevents the RIS-assisted RAN from becoming a cumbersome monolithic system. Furthermore, IRaaS creates opportunities for vendors of intelligent reflection services and balances the interests of both telecom operators and property owners. In summary, IRaaS is expected to accelerate the integration of RIS into mobile networks, fostering an authentic smart radio environment for future mobile communications.

{
It is worth noting that progressively enhancing the performance and impact of IRaaS will require coordinated efforts across multiple dimensions in the future. First, as hardware fundamentally determines the performance ceiling of IRaaS, future RIS designs must continue to improve key hardware capabilities, e.g., reflection efficiency and switching delay. Second, on the software side, model-free configuration methods must ensure compatibility across different generations of RIS hardware during incremental evolution. As RIS hardware capabilities scale up (e.g., in terms of quantity or per-unit performance), model-free algorithms should be able to fully unleash their potential. Third, the degree of integration between RIS and the CN, particularly whether RCF should be realized as an AF or as an NF, remains an open question that merits further exploration. }

\bibliographystyle{IEEEtran}
\bibliography{bibfile}

@article{basar2019wireless,
  title={Wireless communications through reconfigurable intelligent surfaces},
  author={Basar, Ertugrul and Di Renzo, Marco and De Rosny, Julien and Debbah, Merouane and Alouini, Mohamed-Slim and Zhang, Rui},
  journal={IEEE Access},
  volume={7},
  pages={116753--116773},
  year={2019},
  publisher={IEEE}
}

@article{zeydan2022service,
  title={Service based virtual {RAN} architecture for next generation cellular systems},
  author={Zeydan, Engin and Mangues-Bafalluy, Josep and Baranda, Jorge and Requena, Manuel and Turk, Yekta},
  journal={IEEE Access},
  volume={10},
  pages={9455--9470},
  year={2022},
  publisher={IEEE}
}

@TechReport{etsiTransRIS, 
author = {ETSI}, 
day = {20}, 
institution = {ETSI}, 
month = {Aug.}, 
note = {V1.1.1}, 
title = {{Reconfigurable Intelligent Surfaces {(RIS)}; {Technological} challenges, architecture and impact on standardization}}, 
type = {Group Report}, 
year = {2023}
}

@article{qadir2016resource,
  title={Resource Pooling for Wireless Networks: Solutions for the Developing World},
  author={Qadir, Junaid and Sathiaseelan, Arjuna and Wang, Liang and Crowcroft, Jon},
  journal={ACM SIGCOMM Comp. Commun. Rev.},
  volume={46},
  number={4},
  pages={30--35},
  year={2016},
  publisher={ACM New York, NY, USA}
}

@ARTICLE{CEBF,
  author={Nadeem, Qurrat-Ul-Ain and Alwazani, Hibatallah and Kammoun, Abla and Chaaban, Anas and Debbah, Mérouane and Alouini, Mohamed-Slim},
  journal={IEEE Open J. Commun. Soc.}, 
  title={Intelligent Reflecting Surface-Assisted Multi-User {MISO} Communication: Channel Estimation and Beamforming Design}, 
  year={2020},
  volume={1},
  number={},
  pages={661-680},
  doi={10.1109/OJCOMS.2020.2992791}}

@ARTICLE{ModelFree,
  author={Wang, Wei and Zhang, Wei and Xiong, Hongkai},
  journal={IEEE Wireless Commun.}, 
  title={Model-Free Configuration of Intelligent Reflecting Surfaces: Toward Pervasive Adaptability and Enhanced Robustness}, 
  year={2024},
  volume={31},
  number={2},
  pages={142-148},
  doi={10.1109/MWC.013.2200416}}

@article{mei2022intelligent,
  title={Intelligent reflecting surface-aided wireless networks: From single-reflection to multireflection design and optimization},
  author={Mei, Weidong and Zheng, Beixiong and You, Changsheng and Zhang, Rui},
  journal={Proc. IEEE},
  volume={110},
  number={9},
  pages={1380--1400},
  year={2022},
  publisher={IEEE}
}

@ARTICLE{C-RAN,
  author={Wu, Jun and Zhang, Zhifeng and Hong, Yu and Wen, Yonggang},
  journal={IEEE Network},
  title={Cloud radio access network {(C-RAN)}: a primer},
  year={2015},
  volume={29},
  number={1},
  pages={35-41},
  doi={10.1109/MNET.2015.7018201}}

@article{shen2021holistic,
 title={Holistic Network Virtualization and Pervasive Network Intelligence for {6G}},
 author={Shen, Xuemin and Gao, Jie and Wu, Wen and Li, Mushu and Zhou, Conghao and Zhuang, Weihua},
 journal={IEEE Commun. Surveys Tuts.},
 volume={24},
 number={1},
 pages={1--30},
 year={1st. Quart. 2022}
}

@article{wang2022intelligent,
  author={Wang, Wei and Zhang, Wei},
  journal={IEEE J. Sel. Areas Commun.},
  title={Intelligent Reflecting Surface Configurations for Smart Radio Using Deep Reinforcement Learning},
  year={2022},
  volume={40},
  number={8},
  pages={1-1},
  doi={10.1109/JSAC.2022.3180787}}

@article{larson2019derivative,
  title={Derivative-free optimization methods},
  author={Larson, Jeffrey and Menickelly, Matt and Wild, Stefan M},
  journal={Acta Numerica},
  volume={28},
  pages={287--404},
  year={2019},
  publisher={Cambridge University Press}
}

@TechReport{gpp38413, 
author = {3GPP}, 
day = {20}, 
institution = {{3rd Generation Partnership Project (3GPP)}}, 
month = {Jul.}, 
note = {V15.0.0}, 
number = {38.413}, 
title = {{5G; NG-RAN; NG Application Protocol (NGAP)}}, 
type = {Technical Specification (TS)}, 
year = {2018}
}

@article{abeywickrama2020intelligent,
  title={Intelligent reflecting surface: Practical phase shift model and beamforming optimization},
  author={Abeywickrama, Samith and Zhang, Rui and Wu, Qingqing and Yuen, Chau},
  journal={IEEE Trans. Commun.},
  volume={68},
  number={9},
  pages={5849--5863},
  year={2020},
  publisher={IEEE}
}

@INPROCEEDINGS{RISintegration,
  author={Katsalis, Kostas and Arfaoui, Afaf and Liaskos, Christos and Triay, Joan},
  booktitle={2024 IEEE Conference on Standards for Communications and Networking (CSCN)}, 
  title={{RIS} Technology Integration with the {5G} System: Challenges and Open Questions}, 
  year={2024},
  volume={},
  number={},
  pages={14-19},
  doi={10.1109/CSCN63874.2024.10849752}}

@ARTICLE{PMvirtual,
  author={Liaskos, Christos and Katsalis, Kostas and Triay, Joan and Schmid, Stefan},
  journal={IEEE Commun. Mag.}, 
  title={Resource Management for Programmable Metasurfaces: Concept, Prospects and Challenges}, 
  year={2023},
  volume={61},
  number={11},
  pages={208-214},
  doi={10.1109/MCOM.002.2300012}}

\end{document}